\documentclass[amsmath,amssymb,
showpacs,twocolumn, nobibnotes,aps,prd]{revtex4-1}

\usepackage{graphicx,amsmath,amssymb}
\usepackage{mathrsfs,eufrak}
\usepackage{dcolumn,color}
\usepackage{bm}
\usepackage{bbold}
\newcommand{\be}{\begin{equation}}
\newcommand{\ee}{\end{equation}}
\def\ba{\begin{aligned}}
\def\ea{\end{aligned}}

\newcommand{\bea}{\begin{eqnarray}}
\newcommand{\eea}{\end{eqnarray}}

\renewcommand{\hat}[1]{{\widehat #1}}

\begin{document}

\title{Two-spin entanglement induced by scattering of backscattering-free chiral electrons in a chern insulator}
\author{M. Soltani$^1$}  
\author{M. Amini$^1$}
\email{msn.amini@sci.ui.ac.ir}
\affiliation{$^1$ Department of Physics, Faculty of Sciences, University
of Isfahan, Isfahan 81746-73441, Iran}

\begin{abstract}
The existence of robust chiral edge states in a finite topologically nontrivial chern insulator is a consequence of the bulk-boundary correspondence. In this paper, we present a theoretical framework based on lattice Green's function to study the scattering of such chiral edge electrons by a single localized impurity.
To this end, in the first step, we consider the standard topological Haldane model on a honeycomb lattice with strip geometry. 
We obtain analytical expressions for the wave functions and their corresponding energy dispersion of the low-energy chiral states localized at the edge of the ribbon.
Then, we employ the $T$-matrix Lippmann-Schwinger approach to explicitly show the robustness of chiral edge states against the impurity scattering. 
This backscattering-free process has an interesting property that the transmitted wave function acquires an additional phase factor.
Although this additional phase factor does not affect quantum transport through the chiral channel it can carry quantum information. 
As an example of such quantum information transport, we investigate the entanglement of two magnetic impurities in a chern insulator through the dissipation-less scattering of chiral electrons.     
\end{abstract}
\maketitle
\section{Introduction~\label{Sec01}}
In the past decade or so, topological materials, due to their exceptional physical properties, have attracted lots of interests especially in the fields of condensed matter and atomic physics~\cite{HasanRMP,ZhangRMP}. 
These kinds of materials exhibit some exotic phenomena such as
dissipation-less edge transport or Majorana fermions.
The existence of such topologically protected edge modes is a consequence of the bulk-boundary correspondence~\cite{HasanRMP,ZhangRMP}, 
which is the existence of $(d-1)$-dimensional boundary sates inside the energy gap of a $d$-dimensional topologically non-trivial band structure. 
The emergence of such robust boundary states in topological materials plays an important role in designing quantum devices without decoherence~\cite{NayakRMP}.
It is important to note that the physical properties of these edge states depend on the topological classes of the materials forming the boundary.
For instance, in the Chern insulators the most convenient way of classifying topological invariant in the Hamiltonian is the  so-called Chern number~\cite{Thouless}, $C$, which shows the existence of $C$ chiral
conducting edge channels~\cite{Laughlin} in which electron is allowed to move only in a specific direction.

In 1988, Duncan Haldane proposed 
a prototype lattice model of spinless electrons which exhibit anomalous quantum  Hall (AQH) effect in a periodic magnetic field such that the resultant flux through the unit cell is zero~\cite{Haldane1988}. 
Indeed, this is a pioneering example of a system with quantized Hall conductance in which the time-reversal symmetry (TRS) breaks in the absence of a net magnetic field or discrete Landau levels.
The phase diagram of this model, as it is known~\cite{Haldane1988}, contains both the trivial and the Chern insulator domains where
the corresponding Chern number of the topological phase is $C = \pm1$ in comparison to the trivial phase in which $C=0$. 
Therefore, this system supports a single chiral edge channel which is protected topologically.
Then, one concludes that such chiral edge electrons are robust
against impurity scattering or disorder and cannot undergo backscattering.
Although this robustness against the weak local perturbations is a consequence of topological protection and has been studied numerically~\cite{Castro,Bernevig,Malki}, there is still a general lack of a rigorous and direct analytical analysis for such dissipation-less transport.  
This is due to the lack of our knowledge about the analytical expressions for the wave functions of topological edge states.
This issue is addressed for some topological systems very recently ~\cite{Duncan}, but it is still an open question for the Haldane model and will be discussed in the following.

Even though Chern insulators are the first discovered type of topological insulators, the experimental realization of such systems in realistic materials succeeded recently~\cite{exp2013}.
Moreover, using the ultracold fermionic atoms in a periodically modulated optical lattice, it is now possible to implement the Haldane model in an experimental setup which allows tunability of the physical properties of the model~\cite{Haldane-exp1,Haldane-exp2}.

On the other hand, one of the 
most interesting aspects of quantum science are
to understand and control the transport of quantum information 
and create entanglement among different subsystems of a
quantum-scale system~\cite{Nielsen}. 
In this regard, one dimensional (1D) dissipation-less channels in topological materials, as we discussed before,  are
promising candidates~\cite{Zoller1,Zoller2,Nori} in which
the spin of an electron
propagating along the 1D edge could act as a flying qubit~\cite{Habgood}.
Although it is obvious that, due to the topological protection of edge states, the scattering of such chiral edge electron by an impurity will not change the electronic conductance (transmission) of the corresponding edge channel, it is still not clear whether it can change other quantum properties of the incident electron or not.
Indeed, the interaction of single impurity with the chiral edge states
may give rise to the generation of an additional phase factor in the chiral electron's wave function. 
This additional phase factor is, however, quite interesting since it can be used as a resource for quantum entanglement generation which is the most basic type of quantum information processing.

The present paper is an attempt to answer the above-raised questions analytically using the scattering phenomena. 
In this paper, we propose a scheme for efficiently generating entanglement between two magnetic impurities (spin-$1/2$) embedded in a Chern insulator.
We consider the Haldane model for a strip geometry (zigzag ribbon) and obtain the corresponding wave functions of the chiral edge electrons analytically.
Using this wave function, we present in detail, the scattering formalism of these chiral edge electrons by the quantum-dot spin qubits fixed at the edge of the zigzag nanoribbon.  
We explicitly, calculate the transmitted wave function of backscattering-free edge electrons scattered by such impurities using the Lippmann–Schwinger equation and Green's function approach~\cite{Economou1990}. 
In this case, the scattered wave function through the localized scatterers acquires a phase factor that we use to entangle two magnetic impurities located at the edge sites of the ribbon as qubits participating in the scattering process.  


The rest of the paper is organized as follows. 
After the Introduction, in section~\ref{Sec02} we consider the general Hamiltonian of the Haldane model to derive the localized wave function of chiral edge states as well as the corresponding lattice Green's function for such states.
We also calculate the scattering of such chiral edge electrons by an impurity which is modeled as a diagonal on-site potential using the standard $T$-matrix approach on a lattice, adopted by many
authors~\cite{Peres,Katsnelson,Bena,Amini1}.
In Section~\ref{Sec03} we perform explicit calculations of the entanglement generated between a flying chiral qubit and a fixed localized spin qubit.
The generalization of this formalism to the case of two spin qubits localized far from each other is also discussed in this section. 
Finally, the conclusions are summarized in section~\ref{Sec04}.

\begin{figure}[t!]
	\center{\includegraphics[width=0.8\linewidth]{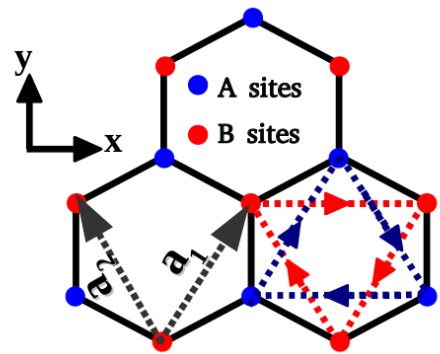}} 
	\caption{The 2D honeycomb lattice structure with two sub-lattices A and B marked by blue and red points respectively and the elementary translation vectors $a_1$ and $a_2$. Dashed arrows show the directions along which the phase factor $\phi$ of the next nearest neighbor hoppings, $t_2$,  is positive, otherwise, $\phi$ is negative.
}
	\label{fig1}
\end{figure}

\section{Haldane model, edge states and Green's function\label{Sec02}}

\begin{figure}[t!]
	\center{\includegraphics[width=0.8\linewidth]{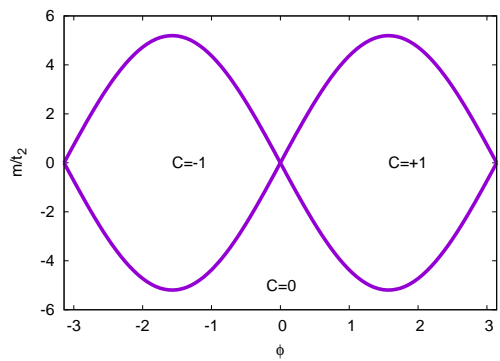}} 
	\caption{The exact phase diagram -as calculated by Haldane~\cite{Haldane1988}- of the topological Haldane model as a function of the on-site energy, $m$ (measured in units of $t_2$) and the staggered phase $\phi$. The regions with non-zero chern number, $C=\pm 1$ are the topologically nontrivial phase which is separated from the trivial phase with zero chern number, $C=0$.
	}
	\label{fig2}
\end{figure}

As a generic example of a Chern topological insulator,
we consider the standard Haldane model~\cite{Haldane1988} for a two-dimensional (2D) honeycomb lattice structure which is illustrated in Fig.(~\ref{fig1}).
This lattice consists of two triangular sub-lattices "A" and "B" indicated by "blue" and "red" sites respectively.
In this model, the motion of spinless fermions at each lattice sites can be described by hopping to nearest- and next-to-nearest-neighbor 
sites.
The tight-binding Hamiltonian for this model
is given by
\bea
&H=&H_0 + H_1, \nonumber \\
&H_0=&\sum_{\langle i,j\rangle}  t_1 (c^\dagger_{iA}c_{jB} + h.c) +m\sum_{i\in A} c^\dagger_{iA}c_{iA}-m\sum_{i\in B} c^\dagger_{iB}c_{iB}, \nonumber\\
&H_1=&\sum_{\langle\langle i,j\rangle\rangle} t_2 e^{i\phi_{ij}} (c^\dagger_{iA}c_{jA} + c^\dagger_{iB}c_{jB}+ h.c), 
\label{H_Haldane}
\eea
where $c^\dagger_i (c_i)$ is the fermionic creation (annihilation) operator at site $i$ and $t_1$ and $t_2$ are real hopping amplitudes between nearest neighbors on the different and the
same sub-lattices, respectively.
The phase factor $\phi_{ij}=\phi$ shown in Fig.~\ref{fig1} is considered to be positive for anticlockwise hoppings and negative for clockwise hoppings. 
Therefore it breaks the time-reversal symmetry of the Hamiltonian since it acts as a staggered magnetic field. Furthermore, 
the on-site energy $+m$ on $A$
sites and $-m$ on $B$ sites  breaks the spatial inversion symmetry of the
model and opens a trivial gap at
the so-called Dirac points.

The well-known topological phase diagram of the Haldane model~\cite{Haldane1988} on a bulk
honeycomb lattice is shown in Fig.~\ref{fig2}. 
The topological phase transition from a chern to trivial insulator takes place at $|m|=3\sqrt{3}t_2\sin{\phi}$. 
The resulting phase regions are indicated by different chern numbers $C=\pm 1$ and $C=0$ respectively in Fig.~\ref{fig2}.
It is important to note that, according to the bulk-boundary correspondence~\cite{HasanRMP,ZhangRMP}, when the Chern number of the bulk system is nonzero, $C=\pm 1$, the boundary
of the confined system provides chiral conducting edge modes.


\subsection{Edge states}

\begin{figure}[t!]
	
	\center{\includegraphics[width=1.1\linewidth]{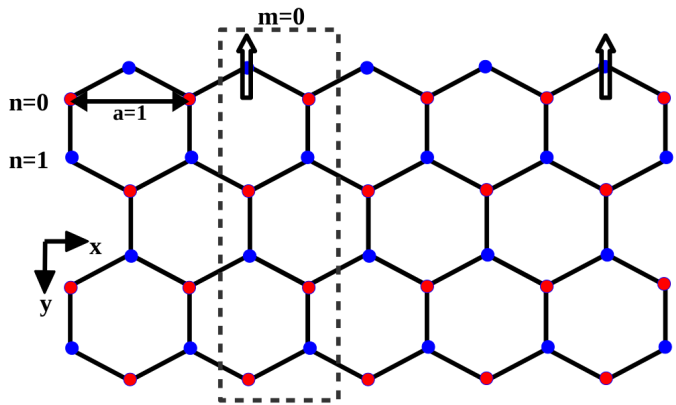}} 
	\caption{Schematic representation of a zigzag honeycomb ribbon with $N_z=4$ zigzag chains across the width of the ribbon.
The dashed rectangle across the width of the ribbon denotes the supercell along this ribbon.
Two typical magnetic impurities located at the top edge of the ribbon are shown at sites $(m=0,n=0,A)$ and $(m=3,n=0,A)$.}
	\label{fig3}
\end{figure}

In the following, we consider the Haldane model for a system with the geometry of a ribbon which has finite height in the $y$-direction whereas is periodic in the $x$−direction with a zigzag edge as depicted in Fig.~\ref{fig3}. 
The ribbon width $N_z$ is defined by the number of zigzag chains along the $y$−direction.
As it is shown in Fig.(~\ref{fig3}), each lattice site of this ribbon can be described simply by a set of three parameters $(m,n,\nu)$ in which $m$ and $n$ are the supercell and zigzag chain index respectively and $\nu = A,B$ refers to the sublattices.
In what follows, the distance of two nearest-neighboring lattice sites on the same sublattice is set as the unit of the length scale, $a=1$.

In the absence of impurity, due to the existence of translational symmetry in the $x$-direction, the wavenumber $k_x$ is still a good quantum number.
We are interested in the topologically nontrivial phase, $C=\pm 1$, in which there is a midgap band crossing (between the valance and the conduction bands) at $k_x = 0$.
This results in the formation of edge bands within the gap of the bulk Haldane model.
Therefore, we can do our calculations for the case $\phi=\pi/2$ to have a large gap and set $t_2=0.3t_1$ to obtain rather flat edge bands.
We also take $m=0$ for simplicity.

\begin{figure}[t!]
	
	\center{\includegraphics[width=1.1\linewidth]{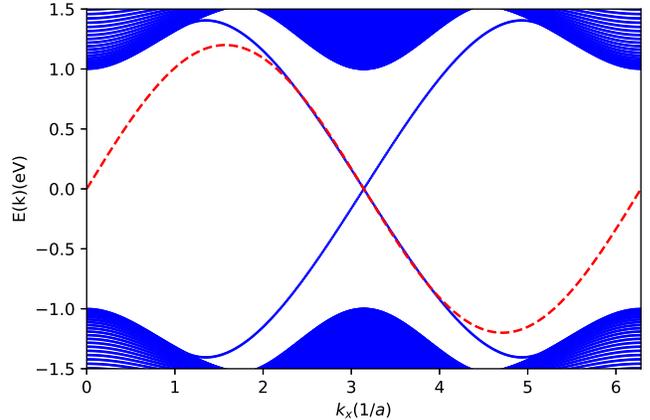}} 
	\caption{Band structure of the Haldane model for a zigzag ribbon of width $N_z=80$ and $t_1=1.0(eV)$, $t_2=0.3(eV)$, and $\phi=\pi/2$
obtained numerically via exact diagonalization. 
The dispersion of the edge states is obvious in the gap region
between the lower band edge of the upper continuum and the upper band edge of the lower continuum.
The dashed (red) lines show a plot of the analytical expression for edge states energy dispersion which is obtained in Eq.~(\ref{E0}).
	}
	
	\label{fig4}
\end{figure}

Fig.~\ref{fig4} shows the band structure of the Haldane
model for a ribbon of width $N_z=80$ which is obtained numerically.
This seems to be sufficiently wide to suppress the backscattering of edge modes~\cite{Malki,Colomes}. 
It is obvious that close to the zero energy, two edge bands composed of edge states cross the Fermi energy.
Such edge states are well known as chiral edge states because they only propagate in one direction along the edge.
These states are robust to any kind of disorder and impurity because there are no states possible for backscattering.
This is what we expect from topological protection and has been investigated numerically~\cite{Malki}. 
We are now going to achieve it analytically in the remainder of this section.

In order to proceed further, we need to have the explicit analytical expressions of the edge states wave-functions as well as their corresponding energy dispersion.
We are particularly interested in the zero-energy solutions of the Schrödinger's equation for the  Hamiltonian of Eq.~(\ref{H_Haldane}), $H|\Psi_{k_x}\rangle=0$.
Since the hopping amplitude $t_2$ (in $H_1$) is smaller than its corresponding hopping $t_1$ (in $H_0$) we can ignor $H_1$ first and take it into account later perturbatively. 
In this regime where the Hamiltonian of Eq.~(\ref{H_Haldane}) only includes hopping between nearest neighbors
the corresponding wave function of the zero energy states has none zero amplitude only in one sublattice (namely "A")~\cite{Amini1,Amini2,Wakabayashi}.
Furthermore, due to the existence of translation symmetry in the $x$ direction, the wave-function of an edge state which is localized at the top edge of the ribbon can be described by the following ansatz
\be
|\Psi_{k_x}(m,n,A)\rangle \propto e^{ik_xm}\psi(n) |m,n,A\rangle.
\label{Psi0}
\ee
Therefore, the equation characterizing the wave function amplitude at the $m=0$ zigzag chain,  for the zero-energy edge states, is  
\be
\psi(0)(e^{i\frac{k_x}{2}}+e^{-i\frac{k_x}{2}})+\psi(1)=0,
\ee
which can be solved as
\be
\psi(1)=2\psi(0)\cos(\frac{k_x}{2}).
\ee
It is now possible to apply the same argument for the wave function amplitude at the $m=1$ zigzag chain
to obtain the same form of the solution.
Thus the wave function amplitude at the sites of the $n$th zigzag chain from the edge of the ribbon is proportional to the factor of 
$\psi(n)=(2\cos(\frac{k_x}{2}))^n$.
We should note that this solution is valid only when $|\cos(\frac{k_x}{2})|<\frac12$ which 
defines the region $\pi-\frac\pi3<k_x<\pi+\frac\pi3$ in which the edge band exists.
Thus, we can now write the normalized edge states as
\be
|\Psi_{k_x}\rangle =\frac{1}{\sqrt{\pi}}\sum_{m,n}  e^{ik_xm} \gamma_{k_x} \psi(n) |m,n,A\rangle, \hspace{2mm} \frac{2\pi}{3}<k_x<\frac{4\pi}{3}
\label{Psi1}
\ee
in which $\gamma_{k_x}=\sqrt{1-4\cos^2(\frac{k_x}{2})}$ is the normalization factor.

Using the wave-function of Eq.~(\ref{Psi1})
and employing the standard first-order perturbation theory, one can now easily obtain the energy dispersion relation of such edge modes. It is the expectation value of the second term of the Hamiltonian in Eq.~(\ref{H_Haldane})  which can be written as   
\be
E(k)=\langle \Psi_{k_x}|H_1|\Psi_{k_x} \rangle =2t_2 \sin{(k_x)}.
\label{E0}
\ee 
The leading order for this energy dispersion of the edge modes reads as 
\be
E(k) \approx 4t_2 k = \hbar v_f k,
\label{E_k}
\ee
where $v_f\approx 4t_2/\hbar$ is the Fermi velocity of chiral edge electrons.

It is now interesting to compare this analytical expression of the energy dispersion obtained for the edge states with the numerical band structure of the Haldane ribbon in Fig.~\ref{fig3}.
The dashed (red) lines in Fig.~\ref{fig3} shows the analytical representation of expression in Eq.~(\ref{E0}) which shows an excellent agreement with the corresponding numerical edge band in the low-energy regime.

It is, however, important to note that within this approximation so long as we are away from the energies where the edge modes enter the continuum of the bulk states, the edge state wave function will not change.
But, if we are interested in the high-energy regime (close to the lower band edge of the upper continuum), we need to consider higher-order corrections. In what follows we restrict our attention to the low-energy regime only.

\subsection{Robustness of chiral edge states against the impurity scattering}
In this subsection we use the Lippmann-Schwinger approach~\cite{Economou1990} to study the effect of
a single impurity on the electronic transmission of chiral edge states in the topologically nontrivial phase of Haldane model.
For simplicity, we drop the sublattice indices from now on since we want to focus on a single chiral edge channel which is composed of edge states localized at the top edge of the ribbon.

For analytical calculations, it is convenient to model a single impurity located at the edge of the ribbon as an on-site potential term, namely,
\be
\hat{V}=V_0 |0,0\rangle\langle 0,0|.
\label{V0}
\ee
To study the effect of this local impurity on the transport properties of chiral electrons in a quantum regime, we need to find the so-called $\hat{T}$-matrix~\cite{Peres,Katsnelson,Bena,Amini1} which is generally written as $\hat{T}=\hat{V}(1-\hat{V}\hat{G})^{-1}$ where $\hat{G}$ is the Green's function of defect-free chiral electrons described by the Hamiltonian of Eq.~(\ref{H_Haldane}). 
Using the impurity potential of Eq.(~\ref{V0}) the $\hat{T}$-matrix reduces to
\be
\hat{T}=\frac{V_0 |0,0\rangle\langle 0,0|}{1-V_0G_{0,0;0,0}}
\ee
where
$G_{0,0;0,0}=\langle 0,0 |\hat{G}|0,0\rangle$ is the matrix element of Green's function operator $\hat{G}$ of the chiral edge states which can be written as
\be
\hat{G}=\int \frac{|\Psi_{k_x}\rangle\langle \Psi_{k_x}|}{E-E(k)+i0^+} dk.
\label{green}
\ee
We now insert into the above formula the general expression of the chiral edge modes as well as their energy dispersion for the low-energy regime which is given by 
Eqs.~(\ref{Psi1}) and ~(\ref{E_k}) respectively. 
The integration in Eq.~(\ref{green}) can be performed according to the methods of residues~\cite{Amini1,Amini2} which results in 
\be
\langle 0 0 |\hat{G}| 0 0 \rangle = i|\psi(0)|^2 v_f. 
\ee

By the same token, we can now derive the following matrix elements which we need later on,
\be
\langle n^\prime m^\prime |\hat{G}| n m \rangle = \int \frac{\psi^*(n^\prime)\psi(n)e^{ik(m-m^\prime)}}{E-E(k)+i0^+} dk.
\ee
We need now to be careful using the methods of residues.
The reason is that when $m>m^\prime (m<m^\prime)$, we need to close the integration contour in the upper (lower) half complex plane.
This results in
\be
\langle n^\prime m^\prime |\hat{G}| n m \rangle = \begin{cases}
    2 i \psi^*(n)\psi(n^\prime)e^{ik(m-m^\prime)}       & \quad \text{if } m>m^\prime\\
    0 & \quad \text{if } m<m^\prime
  \end{cases}.
\ee

Having made these expressions for the matrix elements of the Green's function,  it is straightforward to calculate  an eigenstate, $|\Psi_{out}\rangle$, of the full Hamiltonian (in presence of the impurity potential) using the Lippmann-Schwinger equation\cite{Economou1990} as
\be
|\Psi_{out}\rangle=|\Psi_{in}\rangle+\hat{G}\hat{T}|\Psi_{in}\rangle=|\Psi_{in}\rangle+\frac{\hat{G}\hat{V}}{1-i\alpha}|\Psi_{in}\rangle
\ee
in which $\alpha=\psi^2(0)V_0$.
If we set the $|\Psi_{in}\rangle=|\Psi_k\rangle$ we get
\be
|\Psi_{out}\rangle = (1+\frac{2i\alpha}{1-i\alpha})|\Psi_k\rangle = e^{i\phi_0} |\Psi_k\rangle.
\label{psiout}
\ee
where $\tan(\phi_0/2)=\alpha$.
In the final expression of Eq.~(\ref{psiout}), the transmission coefficient is always one and the reflection coefficient is always zero.   This result illustrates that the propagating chiral modes at the edge of the ribbon will not backward scatter
and will totally transmit through the impurity potential.   
It comes as no surprise since we expected the chiral edge states to be robust against the impurity scattering due to the topological protection (this was also shown by numerical computations in Ref.~\cite{Malki}).

The interesting thing in Eq.~(\ref{psiout}) is the appearance of an additional phase $\phi_0$ in the transmitted wave function with respect to the incident wave function. 
Although this additional phase factor can not affect the transmission properties of the chiral edge channel, it can be used to entangle two different magnetic impurities located at the edge of the ribbon. This is what we will discuss in the next section.

\section{Entanglement Analysis \label{Sec03}}
In this section, we study the phenomenon of quantum entanglement generation between two magnetic impurities through the electron scattering which is discussed in Ref.~\cite{PRL-Bose}.
We present two different cases of single and double impurity in the following and show how the pairwise entanglement generated via chiral electron scattering in a chern insulator is related to the additional phase factor which we introduced before.
\subsection{Single impurity}
Using the results obtained in the previous section and employing the approach of Ref.~\cite{Amini3},  we can now study the entanglement between a single localized magnetic impurity
and low-energy chiral electrons in the Haldane model.
We observe that, in the presence of a single on-site impurity, 
for a given incoming chiral state, the reflection and transmission amplitudes of the final wave state read respectively as
$r=0$ and $t=e^{i\phi_0}$.
Now, let a single on-site spin-$1/2$ magnetic impurity be introduced to the system with the following impurity potential
\be
\hat{V}=V_0\hat{S_1}\cdot\hat{S_2} |00\rangle\langle 00|,
\ee
where $S_1$ and $S_2$ are the corresponding dimensionless spin operators of the incident chiral electron and localized impurity respectively.
It is convenient to work in the computational basis in which the spin interaction $\hat{S_1}\cdot\hat{S_2}$ can be written as
\be
\hat{S_1}\cdot\hat{S_2} = \frac{1}{4}\left(|t\rangle\langle t|-3|s\rangle\langle s|\right),
\ee
where $t$ and $s$ refer to triplet and singlet spin states, respectively.
Combining this equation with Eq.~(\ref{psiout}) one can easily obtain the $\hat{T}$-matrix for the scattered wave state as
\be
\hat{T}=e^{i\phi_{0t}}|t\rangle\langle t|+e^{i\phi_{0s}}|s\rangle\langle s|,
\ee
where $\tan(\phi_{0s})=-\frac{3}{4}V_0|\psi(0)|^2$ and $\tan(\phi_{0t})=\frac{1}{4}V_0|\psi(0)|^2$.
If we consider the spatial and spinorial
spaces in an incoming wave state as
\be
|\Psi_{in} \rangle = |\Psi_{k_x}\rangle | \uparrow \downarrow \rangle 
\ee
which corresponds to spin up incident chiral electron with the wavenumber $k_x$ while the impurity has spin down in the $z$ direction.
It is now straightforward to calculate the outgoing wave state using the Lippmann–Schwinger equation as~\cite{Amini3}
\bea
|\Psi_{out} \rangle &=& |\Psi_{k_x}\rangle (\hat{T}| \uparrow \downarrow \rangle)=e^{i\phi_{0s}} |s\rangle + e^{i\phi_{0t}} |t\rangle \\
&=& e^{i\frac{\phi_{0s}+\phi_{0t}}{2}} \left(\cos(\frac{\Delta\phi}{2}) | \uparrow \downarrow \rangle + i\sin(\frac{\Delta\phi}{2}) | \downarrow \uparrow \rangle\right).\nonumber
\eea
This shows that the outgoing state is an entangled state and for $\Delta\phi=\phi_{0s}-\phi_{0t}=\pi/2$ reaches its maximum value.

\subsection{Double impurity}
In this subsection, the previous problem should be extended to the scattering of a  chiral electron by two separate noninteracting magnetic impurities which are schematically shown in Fig.~\ref{fig3}.
Due to the absence of backscattering from the first impurity which we discussed before (perfect transmission), the 
overall transmission coefficient can be written as
 \be
\hat{t}=\hat{t_2}\hat{t_1}
\ee
where $\hat{t_1}(\hat{t_2})$ is the transmission through the first (second) impurity potential.
It is now convenient to write $\hat{t_1}$ and $(\hat{t_2})$ in terms of the computational basis as
\bea
\hat{t_1} &=& \hat{t} \otimes \mathbb{1}_3\nonumber \\
\hat{t_2} &=& \hat{t} \otimes \mathbb{1}_2
\eea
where $\mathbb{1}_2$ and $\mathbb{1}_3$ are the identity matrices on the spin spaces of first and second impurities respectively where
\be
\hat{t}= \begin{bmatrix}
    e^{i\phi_{0t}}       & 0& 0&  0 \\
    0       & \frac{e^{i\phi_t}+e^{-i\phi_s}}{2} & \frac{e^{i\phi_t}-e^{-i\phi_s}}{2} & 0 \\
    0 & \frac{e^{i\phi_t}-e^{-i\phi_s}}{2}& \frac{e^{i\phi_t}+e^{-i\phi_s}}{2} &0 \\
    0      & 0 & 0 & e^{i\phi_{0t}}
\end{bmatrix}.
\ee
If we now set $|\Psi_{in}\rangle=|\psi_{k_x}\rangle |\uparrow\downarrow\downarrow\rangle$ the outgoing wave state is
\be
\hat{t_2}\hat{t_1} |\uparrow\downarrow\downarrow\rangle,
\ee
and after tracing over the electron spin degree of freedom we can obtain the two-impurity reduced density matrix as
\begin{widetext}
\be
\hat{\rho}= \begin{bmatrix}
    0       & 0& 0&  -ie^{-i\phi_{0s}}\sin{(\frac{\Delta\phi_0}{2})}\cos{(\frac{\Delta\phi_0}{2})} \\
    0       & (\sin{(\frac{\Delta\phi_0}{2})}\cos{(\frac{\Delta\phi_0}{2})})^2 & 0 & 0 \\
    0 & 0 & \sin^2{(\frac{\Delta\phi_0}{2})} &0 \\
    ie^{i\phi_{0s}}\sin{(\frac{\Delta\phi_0}{2})}\cos{(\frac{\Delta\phi_0}{2})}      & 0 & 0 & \cos^2{(\frac{\Delta\phi_0}{2})}.
\end{bmatrix}
\ee
\end{widetext}

Given $\hat{\rho}$ one can calculate the corresponding degree
of the entanglement in the final spin state of the system by means of the negativity~\cite{Peres}.
It is defined as the sum of the absolute values of the eigenvalues $\lambda_i$ of the partially transposed reduced density matrix which results in 
\be
\mathcal{N}=\sum_{\lambda_i<0} |\lambda_i|=-\cos^2{(\frac{\Delta\phi_0}{2})}+\sqrt{\cos^4{(\frac{\Delta\phi_0}{2})}+\sin^2{(\Delta\phi_0)}}.
\ee
This shows that the chiral electron scattering leads to entanglement generation between the spin impurities  and the maximal entanglement is reached at $\Delta\phi_0=\frac{\pi}{2}$ which results in $\mathcal{N}=\frac{\sqrt{5}-1}{2}$.

\section{Summary \label{Sec04}}
In summary, in this work, we have introduced a scheme for entangling
two distant magnetic impurities in a chern insulator
through the interaction with a single chiral electron.
We have shown explicitly
and analytically that, when a chiral edge electron scatters by an impurity which is located at the edge 
of the Haldane ribbon acquires an additional phase.
We found that this additional phase can be used to generate entanglement between two magnetic spin-$\frac12$ impurities far away from each other.
Moreover, We have derived analytically and validated numerically the wave functions and corresponding energy spectrum of the chiral edge states that appeared in the Haldane model which can be used in future studies of this model analytically.          

\begin{acknowledgments}
We gratefully acknowledge discussion with R. Fazio during the improvment of this work.
MA also acknowledges the support of the Abdus Salam (ICTP) associateship program.
\end{acknowledgments}



\begin{thebibliography}{}

\bibitem{HasanRMP} M. Z. Hasan and C. L. Kane, Rev. Mod. Phys. 82, 3045 (2010).
\bibitem{ZhangRMP} X.-L. Qi and S.-C. Zhang, Rev. Mod. Phys. 83, 105713 (2011).
\bibitem{NayakRMP} C. Nayak, S. H. Simon, A. Stern, M. Freedman and S. Das Sarma,  Rev. Mod. Phys. 80, 1083 (2008).
\bibitem{Thouless} D. J. Thouless, M. Kohmoto, M. P. Nightingale, and M. den Nijs, Phys. Rev. Lett. 49, 405 (1982). 
\bibitem{Laughlin} R. B. Laughlin, Phys. Rev. B 23, 5632 (1981).
\bibitem{Haldane1988} F. D. M. Haldane, Phys. Rev. Lett. 61, 2015 (1988).
\bibitem{Castro} E. V. Castro, M. P. Lopez-Sancho, and M. A. H. Vozmediano,
Phys. Rev. B 92, 085410 (2015).
\bibitem{Bernevig} E. Prodan, T. L. Hughes, and B. A. Bernevig, Phys. Rev. Lett.
105, 115501 (2010).
\bibitem{Malki} M. Malki and G. S. Uhrig, Phys. Rev. B 95, 235118 (2017).
\bibitem{Duncan} C. W. Duncan, P. Öhberg, and M. Valiente, Phys. Rev. B 97, 195439 (2018).
\bibitem{exp2013} C.-Z. Chang, J. Zhang, X. Feng, J. Shen, Z. Zhang, M. Guo, K.
Li, Y. Ou, P. Wei, L.-L. Wang, Z.-Q. Ji, Y. Feng, S. Ji, X. Chen,
J. Jia, X. Dai, Z. Fang, S.-C. Zhang, K. He, Y. Wang, L. Lu,
X.-C. Ma, and Q.-K. Xue, Science 340, 167 (2013).
\bibitem{Haldane-exp1} G. Jotzu, M. Messer, R. Desbuquois, M. Lebrat, T. Uehlinger, D. Greif, and T. Esslinger, Nature (London) 515, 237 (2014).
\bibitem{Haldane-exp2} Z. Wu, L. Zhang, W. Sun, X.-T. Xu, B.-Z. Wang, S.-C. Ji, Y.
Deng, S. Chen, X.-J. Liu, and J.-W. Pan, Science 354, 83 (2016).
\bibitem{Nielsen} M. A. Nielsen and I. L. Chuang, Quantum Computation and Quantum Information (Cambridge University Press,
Cambridge, UK, 2000).
\bibitem{Zoller1} N. Y. Yao, C. R. Laumann, A. V. Gorshkov, H. Weimer,
L. Jiang, J. I. Cirac, P. Zoller, and M. D. Lukin, Nature Commun. 4, 1585 (2013).
\bibitem{Zoller2} C. Dlaska, B. Vermersch, and P Zoller, Quantum Sci. Technol. 2, 015001 (2017).
\bibitem{Nori} C. Gneiting, D. Leykam, and F. Nori, Phys. Rev. Lett. 122, 066601 (2019).
\bibitem{Habgood} M. Habgood, J. H. Jefferson, A. Ramšak, D. G. Pettifor, and G. A. D. Briggs, Phys. Rev. B 77, 075337 (2008).
\bibitem{Economou1990} E. N. Economou, \emph{Green's function in quantum physics}, Springer Verlag, Berlin (2006).
\bibitem{Peres} N. M. R. Peres, F. Guinea, and A. H. Castro Neto, Phys. Rev. B 73, 125411 (2006).
\bibitem{Katsnelson} T. O. Wehling, A. V. Balatsky, M. I. Katsnelson, A. I. Lichtenstein, K. Scharnberg, and R. Wiesendanger,
Phys. Rev. B 75, 125425 (2007).
\bibitem{Bena} C. Bena, Phys. Rev. Lett. 100, 076601 (2008).
\bibitem{Amini1} M. Amini, M. Soltani and M. Sharbafiun, Physical Review B 99 (8), 085403 (2019).
\bibitem{Colomes} E. Colomés and M. Franz, Phys. Rev. Lett. 120, 086603 (2018).
\bibitem{Amini2} M. Amini and M. Soltani, Journal of Physics: Condensed Matter 31 (21), 215301 (2019).
\bibitem{Wakabayashi} H. Y. Deng and K. Wakabayashi, Phys. Rev. B 90, 115413 (2014).
\bibitem{PRL-Bose} A. T. Costa, Jr., S. Bose, and Y. Omar, Phys. Rev. Lett. 96, 230501 (2006).
\bibitem{Amini3} M. Amini, M. Soltani, E. Ghanbari-Adivi and M. Sharbafiun, Quantum Inf Process 18 (3), 78 (2019).
\bibitem{Peres} A. Peres, Phys. Rev. Lett. 77, 1413 (1996).


















\end{thebibliography}
\end{document}